\documentstyle[12pt,epsfig]{article}
\begin{document}
\setlength{\baselineskip}{0.75cm}
\setlength{\parskip}{0.45cm}
\begin{titlepage}
\begin{flushright}
DO-TH 97/01 \linebreak
January 1997
\end{flushright}
\vskip 0.8in
\begin{center}
{\Large\bf Detailed Next-to-Leading Order Analysis of Deep Inelastic
Neutrino Induced Charm Production off Strange Sea Partons}
\vskip 0.5in
{\large M.\ Gl\"{u}ck, S.\ Kretzer and E.\ Reya}
\end{center}
\vskip .3in
\begin{center}
{\large Institut f\"{u}r Physik, Universit\"{a}t Dortmund \\
D-44221 Dortmund, Germany}
\end{center}
\vskip 0.5in
{\large{\underline{Abstract}}}

\noindent
Neutrino induced deep inelastic charm production off strange sea partons
in the nucleon
is studied in the framework of the $\overline{\rm{{MS}}}$ fixed flavor
factorization scheme. The momentum ($z$) distributions of the produced
D mesons as calculated in the next--to--leading order QCD analysis are
presented and compared to their leading order counterparts. Perturbative
stability within this formalism is demonstrated and the compatibility of
recent next-to-leading order (NLO) strange quark distributions with
available dimuon data is investigated. 
Our NLO ($\overline{\rm{{MS}}}$) results
provide, for the first time, the tools required for a consistent and
complete NLO ($\overline{\rm{{MS}}}$) analysis of charged 
current dimuon data in
order to extract the strange sea density in NLO.
It is furthermore shown that the
absolute predictions of radiatively (dynamically) generated strange sea
densities are in agreement with all present measurements.
\end{titlepage}
%
%
\noindent
Recent attempts \cite{ref1,ref2} to determine the strange sea content of
the nucleon , $s(x,Q^2)$, from a next-to-leading order (NLO) QCD analysis
of neutrino induced charm production (opposite-sign dimuon events) were
undertaken in the framework of the NLO formalism of \cite{ref3}. In this
formalism one considers, in addition to the leading order (LO) process
$W^+ s \rightarrow c$, etc., the $W^+ g \rightarrow c \bar{s}$ gluon fusion
contribution with $m_s \neq 0$, i.e.\ one employs a finite mass
regularization and subtracts from it that (collinear) part which is already
contained in the renormalized, $Q^2$--dependent $s(x,Q^2)$. In a previous
publication \cite{ref4} we have started a study of NLO QCD effects based on
the conventional $\overline{\rm{{MS}}}$ factorization scheme where the mass
of the strange quark is neglected ($m_s=0$) 
and the corresponding mass singularities
are dimensionally regularized \cite{ref5}. Such an analysis is imperative
in connection with all those NLO parton distributions and 
cross sections given
in this, most frequently utilized, 
massless $\overline{\rm{{MS}}}$ factorization scheme.
The analysis in \cite{ref4} was incomplete, however, in the sense of not
addressing the detailed momentum distributions of the produced c--quark
(D--meson) utilized in \cite{ref1,ref2}.
The present paper is devoted to this
issue and the relevant expressions are now calculated in 
their full differential
form, by extending the original inclusive analysis in \cite{ref5}.

The formulae and notations of \cite{ref4} will be strictly taken over
everywhere, being extended merely by the introduction of the fractional
charm momentum variable  
$\zeta \equiv p_c \cdot p_N / q \cdot p_N$, \linebreak 
i.e.\ $H_i^{q,g}(\xi ',\mu^2 ,\lambda )$ in \cite{ref4} will be replaced by
$H_i^{q,g}(\xi ',\zeta ,\mu^2 ,\lambda )$ where the fermionic and gluonic
NLO coefficient functions $H_{i=1,2,3}^q$ and $H_i^g$ for heavy quark
(charm) production derive from the subprocess $W^+ s \rightarrow c g$
(together with the virtual vertex correction to $W^+ s \rightarrow c$)
and $W^+ g \rightarrow c {\bar{s}}$, respectively. Furthermore,
$d^2\sigma^{\nu ({\bar{\nu}})} / dx dy$ in \cite{ref4} will be replaced by
$d^3\sigma^{\nu ({\bar{\nu}})} / dx dy dz$ 
with the relevant semi--inclusive
structure functions, describing the momentum distributions of the produced
D--mesons, being obtained from a convolution with the charm fragmentation 
function $D_c(z)$, 
{\footnote{Instead of $z \equiv p_D \cdot p_N / q \cdot 
p_N$, it is customary to use {\cite{ref1,ref2,ref7}}
$z_p \equiv p_D / p_D^{max}$ which differs only marginally from $z$ in 
the relevant experimental region 
($z \sim$ \hspace{-0.4cm}\raisebox{1ex}{$>$}
$0.05$).}}

\begin{eqnarray}  \nonumber
{\cal{F}}_i^c(x,z,Q^2) &=& s'(\xi,\mu^2_F)\ D_c(z) \\ \nonumber &+&
\frac{\alpha_s(\mu^2_F)}{2\pi}
\int_{\xi}^1 \frac{d\xi '}{\xi '} 
\int_{max(z,\zeta_{min})}^1 \frac{d\zeta}{\zeta} 
\left[H_i^q(\xi ',\zeta,\mu^2_F,\lambda)\ s'(\frac{\xi}{\xi '},\mu^2_F)
\right. \\ 
&+& \left.  
H_i^g(\xi ',\zeta,\mu^2_F,\lambda)\ g(\frac{\xi}{\xi '},\mu^2_F)
\right] D_c(\frac{z}{\zeta}) \ \ .
\end{eqnarray}
where ${\cal{F}}_1^c \equiv F_1^c$, ${\cal{F}}_3^c \equiv
F_3^c/2$, ${\cal{F}}_2^c \equiv F_2^c/2\xi$ with $\xi=x(1+m_c^2/Q^2)$,
$\zeta_{min} = m_c^2 / {\hat{s}} = (1-\lambda ) \xi ' / (1-\lambda \xi ')$
and $\lambda = Q^2 / (Q^2+m_c^2)$. The $H_i^{q,g}$ 
are given in the Appendix and 
our choice for the factorization scale will be $\mu_F^2 = Q^2+m_c^2$,
although the results are not very sensitive to this specific choice. The 
fractional momentum of the D--meson is denoted by $z \equiv 
p_D \cdot p_N / q \cdot p_N$ and the 
(factorization scale independent) charm
fragmentation function is taken as \cite{ref6}
\begin{equation}
D_c(z) = N \left\{ z \left[ 1-z^{-1}-\varepsilon_c/(1-z)
\right]^2\right\}^{-1}
\end{equation}
with the normalization constant $N$ being related to $\varepsilon_c$ via
$\int_0^1 dz D_c(z) = 1$, while 
$\varepsilon_c$ itself will be fitted to the
measured z--distributions. One can infer these distributions by utilizing
the LO version of eq.\ (1) which has a simple factorized form
\begin{equation}
\frac{d^3\sigma_{LO}^{\nu(\bar{\nu})}}{dx\ dy\ dz}\ =
\frac{d^2\sigma_{LO}^{\nu(\bar{\nu})}}{dx\ dy}\ D_c(z)
\end{equation}
together with the corresponding parameters in the LO analysis \cite{ref7}
of which $\varepsilon_c^{LO} = 0.20 \pm 0.04$ is the most relevant now
\cite{ref8}. It should be mentioned that the CCFR group has also performed
a 'NLO' analysis \cite{ref1} of their dimuon data using the simple 
factorized LO expression in eq.\ (3) instead of the correct (convoluted)
NLO expressions in (1), which are just the finite--mass extensions of the
well known NLO results for light (massless) quarks \cite{ref9}. We shall
analyze the 'data' \cite{ref7} as parametrized in LO by eq.\ (3). 

To simplify the presentation of our results we define \cite{ref4}
\begin{equation}
\xi s(\xi,z,Q^2)_{eff}\ \equiv\ \frac{1}{2}\
\frac{\pi(1+Q^2/M_W^2)^2}{G_F^2M_N E_{\nu}}\ \left|V_{cs}\right|^{-2}\
\frac{d^3\sigma^{(c \bar{s})}}{dx\ dy\ dz}
\end{equation}
which has also been studied experimentally \cite{ref1} and where the
superscipt $c \bar{s}$ refers just to the CKM non--suppressed ($V_{cs}$)
component of 
$s'\ \equiv\  \left|V_{cs}\right|^2 s\ +\ \left|V_{cd}\right|^2
(d+u)/2$
in eq.\ (1). In LO the cross section in (4) reduces to \cite{ref4}, cf.\
eqs.\ (1) and (3), 
\begin{equation}
\xi s(\xi,z,Q^2)_{eff}\ =\ (1-\frac{m_c^2}{2 M_N E_{\nu} \xi})\ \xi
s(\xi,\mu^2)\ D_c(z) \ \ \ .
\end{equation}      
Our LO and NLO results for $s_{eff}$ are shown in figs.\ 1 and 2 using
the GRV \cite{ref10} and CTEQ4 \cite{ref11} parton densities, 
respectively. It should be emphasized that the 'data' on
$d^3 \sigma / dx dy dz$ 
are well reproduced in NLO provided one selects a 
harder nonperturbative charm fragmentation function $D_c(z)$ corresponding
to $\varepsilon_c^{NLO} = 0.06$ in eq.\ (2) as expected to be necessary for
compensating  the softening effects due to the NLO ${\cal{O}}(\alpha_s)$
corrections (as is evident from comparing the dashed and solid curves in
figs.\ 1 and 2). Similar ${\cal{O}}(\alpha_s)$ softening effects due to 
gluon bremsstrahlung in $e^+ e^- \rightarrow D X$ imply 
\cite{ref12,ref13} $\varepsilon_c^{NLO} = 0.06 \pm 0.03$. It is
interesting to note that similar values for $\varepsilon_c^{NLO}$, i.e.\
$\varepsilon_c^{NLO} \simeq 0.06$ are obtained in NLO by comparing the
predictions of the NLO perturbative \underline{fixed} order QCD corrections
to $e^+ e^- \rightarrow D X$, based on the corresponding cross sections in
\cite{ref14}, with the ARGUS data \cite{ref8,ref15} ($\sqrt{s} \simeq
10\ {\rm{GeV}}$). 

Apart from the very large $z$ region 
($z \sim$ \hspace{-0.5cm}\raisebox{1ex}{$>$}
$0.8$), where the perturbative $\ln (1-z)$ singularities have to be 
resummed \cite{ref16} and nonperturbative higher twist contributions 
\cite{ref17} will play a dominant role, our LO and NLO results in figs.\
1 and 2 show a remarkable perturbative stability throughout the 
experimentally relevant $z$ and $x$ region 
($z \sim$ \hspace{-0.5cm}\raisebox{1ex}{$<$}
$0.8$,   
$0.01 \sim$ \hspace{-0.45cm}\raisebox{1ex}{$<$}
$x \sim$ \hspace{-0.45cm}\raisebox{1ex}{$<$}
$0.4$) for both sets of parton densities, despite the fact that we have
(arbitrarily) fixed $m_c = 1.5\ {\rm{GeV}}$ in LO and NLO. It should be
reemphasized that a \underline{proper} utilization of the corresponding
LO and NLO parton densities and coupling constants (and possibly $m_c$)
is essential for the observed perturbative stability of the physical cross
sections in eqs.\ (1), (4) and (5). The results in fig.\ 1 support the
dynamically predicted $s(x,Q^2)$, generated purely radiatively from 
$s(x,\mu^2)=0$ at $\mu^2_{LO,NLO}=0.23,\ 0.34\ {\rm{GeV}}^2$ \cite{ref10}.
On the other hand, the CTEQ4 densities \cite{ref11} overshoot somewhat the
shaded 'data' band due to the larger strange sea densities. 

The observed perturbative stability, as shown for example in fig.\ 1,
can possibly be improved by adopting a different charm mass in LO and NLO,
e.g.\ $m_c=1.3\ {\rm{GeV}}$ and $m_c=1.7\ {\rm{GeV}}$, respectively, 
following refs.\ \cite{ref1,ref2}. The observed variations of the LO and
total NLO results are shown in fig.\ 3.

In order to determine the NLO strange sea density, the CCFR collaboration 
\cite{ref1,ref2} employed, as discussed at the beginning, the NLO
mass--regulated ($m_s\neq 0$) $W^+ g \rightarrow c {\bar{s}}$ gluon fusion
contribution within the ACOT framework \cite{ref3} where the NLO 
${\cal{O}}(\alpha_s)$ quark--initiated contribution
($W^+ s \rightarrow c g$) is \underline{dis}regarded (since it has not 
been calculated yet for $m_s \neq 0$). An explicit (more differential)
calculation of this gluon--initiated production process along the lines of
\cite{ref3} shows \cite{ref18} that $d^3 \sigma / dx dy dz$ becomes 
insensitive to $m_s$ provided one chooses 
$m_s \sim$ \hspace{-0.5cm}\raisebox{1ex}{$<$}$\ 200\ {\rm{MeV}}$,
as is usually done \cite{ref1,ref2}, and practically coincides with our
$\overline{\rm{{MS}}}$ results, e.g.\ the dashed--dotted curves in fig.\ 1.
Moreover, these results depend very little on the specific choice of the
initial gluon density for the experimentally relevant ($x,Q^2$) domain,
as is obvious from the dashed--dotted curves in figs.\ 1 and 2; the gluon
density, obtained in the original CCFR analysis \cite{ref1,ref2},
gives again a very similar result. It should, however, be noted that CCFR
\cite{ref1} approximated the full convolutions of the NLO gluonic 
contribution in eq.\ (1) by a factorized expression similar to the one in 
eq.\ (3) while keeping the LO $D_c(z)$ with 
$\varepsilon_c^{LO}\simeq 0.20$. This 'approximated' factorized expression
yields rather different $z$-- and $x$--distributions. Despite the fact
that such a factorized NLO approach is theoretically not justified, this
inconsistency could be partly 'compensated' by choosing a different value
for $\varepsilon_c$ (and possibly also for $m_c$). In the latter case
the extraction of $s_{NLO}$ becomes rather ambiguous, since the utilized 
value of $\varepsilon_c$ cannot be consistently obtained within an 
analogous NLO analysis of, say, $e^+ e^- \rightarrow D X$. On top of this
comes the contribution from the NLO quark--initiated subprocess 
$W^+ s \rightarrow c g$ and the NLO vertex correction to 
$W^+ s \rightarrow c$, so far neglected by the CCFR analysis \cite{ref1,
ref2}, which is not negligible according to the dotted curves in
figs.\ 1 and 2. It is thus clear that a consistent and complete 
NLO--\underline{re}analysis of the CCFR dimuon data would certainly be
worthwhile. 

For illustration we finally compare the NLO strange sea density inferred
by the CCFR collaboration \cite{ref1} with their LO one in 
fig.\ 4, where the
GRV94 and CTEQ4 strange densities are also shown at $Q^2=4$ and
$22\ {\rm{GeV}}^2$. 
The three different strange sea densities are comparable
in the limited $x$--domain of the CCFR dimuon data at fixed $Q^2$ (as
indicated). Outside this region the enhancement of the 'extracted'
$s_{NLO}(x,Q^2)$ with respect to, say, the GRV $s_{NLO}$ should not be
taken too literally since the extraction beyond the measured region depends
on the assumed input $x$--shape as well as on the 
$Q^2$--evolution.{\footnote{
It should be noted that CCFR has assumed a \underline{constant}
value for 
\begin{eqnarray} \nonumber
\kappa(Q^2) \equiv \int_0^1 \left[ xs(x,Q^2)+x{\bar{s}}(x,Q^2)\right] dx
{\bigg/} \int_0^1 \left[ x{\bar{u}}(x,Q^2)+x{\bar{d}}(x,Q^2)\right] 
dx\ \ \ ,
\end{eqnarray}
using $\kappa(Q^2) \simeq 0.48$. 
This is partly responsible for an enhancement
of $s(x,Q^2)$ at smaller values of $Q^2$, since a LO/NLO QCD 
$Q^2$--evolution gives $\kappa(4\ {\rm{GeV}}^2)/ \kappa(20\ {\rm{GeV}}^2)
\simeq 0.8.$}}
On the other hand the excess of the LO/NLO CTEQ4 predictions over the CCFR 
'data' band (as extracted in LO) in fig.\ 2 is caused by the strong 
enhancement of the CTEQ4 strange quark densities with respect to the 
LO/NLO GRV ones, in particular for  
$x \sim$ \hspace{-0.5cm}\raisebox{1ex}{$>$}$\ 0.03$.

To summarize, we have calculated and analyzed within the 
$\overline{\rm{{MS}}}$ fixed flavor scheme the momentum ($z$) distributions
$d^3 \sigma / dx dy dz$ of the $D$ mesons produced in neutrino induced 
deep inelastic charm production off strange sea partons (opposite--sign
dimuon events) according to the quark-- and gluon--initiated NLO
subprocesses $W^+ s \rightarrow c g$ (together with the vertex correction
to the LO Born process $W^+ s \rightarrow c$) and 
$W^+ g \rightarrow c {\bar{s}}$, respectively. Perturbative LO/NLO 
stability within this formalism has been obtained in the safe \linebreak
$z \sim$ \hspace{-0.5cm}\raisebox{1ex}{$<$}$\ 0.8$ region and the 
compatibility of recent LO and NLO strange quark sea densities with 
available CCFR dimuon data \cite{ref1,ref7} has been investigated.
We studied also the mass--regulated ($m_s\neq 0$) $W^+ g \rightarrow
c {\bar{s}}$ gluon fusion contribution within the ACOT framework 
\cite{ref3}, originally used by CCFR for extracting a NLO strange sea
density \cite{ref1} from their dimuon data, where the NLO 
${\cal{O}}(\alpha_s)$ quark--initiated $W^+ s\rightarrow c g$ contribution
is disregarded. This $m_s\neq 0$ gluon initiated contribution turns out
to be insensitive to $m_s$ provided one chooses, as usual, 
$m_s \sim$ \hspace{-0.5cm}\raisebox{1ex}{$<$}$\ 200\ {\rm{MeV}}$ 
and practically coincides with our $\overline{\rm{{MS}}}$ results
\cite{ref18}. 
Nevertheless the CCFR NLO analysis \cite{ref1} of the dimuon data for
extracting $s_{NLO}(x,Q^2)$ is inconsistent due to the use of a 
theoretically unjustified 'approximated' factorized expression in $x$
and $z$ for $d^3 \sigma^{\nu ({\bar{\nu}})} / dx dy dz$, as well as
incomplete due to the neglect of the quark--initiated ${\cal{O}}(\alpha_s)$
contribution. A consistent and complete NLO--\underline{re}analysis of
the CCFR dimuon data, using the fully differential and complete 
NLO ($\overline{\rm{{MS}}}$) contributions as presented in this article, 
would therefore be worthwhile.
%
\section*{Acknowledgements}
This work has been supported in part by the
'Bundesministerium f\"{u}r Bildung, Wissenschaft, Forschung und
Technologie', Bonn.
\newpage
\setcounter{equation}{0}
\def\theequation{A\arabic{equation}}
\section*{Appendix}
The fermionic NLO ($\overline{\rm{{MS}}}$) coefficient functions
$H_i^q$ for heavy quark (charm) production in eq.\ (1), calculated from
the subprocess $W^+ s \rightarrow g c$ and the vertex correction to 
$W^+ s \rightarrow c$ are given by
\begin{eqnarray} \nonumber
H_1^q(\xi,\zeta,\mu^2_F,\lambda) &=&
\delta (1-\zeta )\  \left\{\  P_{qq}^{(0)}(\xi)  \ln 
\frac{Q^2+m_c^2}{\mu^2_F}  \right.  \\ \nonumber &+& \left.
\frac{4}{3} \left[ 1-\xi + (1-\xi) \ln \frac{(1-\xi)^2}{\xi(1-\lambda
\xi)} - 2 \xi \frac{\ln \xi}{1-\xi} + 2 \xi 
\left( \frac{1}{1-\xi} \ln \frac{(1-\xi)^2}{1-\lambda \xi} \right)_+
\right] \right\} \\ \nonumber
&+& \frac{4}{3} \left\{ - \delta(1-\xi) \delta(1-\zeta) \left[
\frac{1}{2} \left( \frac{1+3 \lambda}{\lambda}\ K_A
+\frac{1}{\lambda} \right)\ +\ 4\ +\ 
\frac{\pi^2}{3} \right] \right. \\ \nonumber
&+& \frac{1-\xi}{(1-\zeta)_{\oplus}}\ +\ (1-\zeta)\left(
\frac{1-\lambda \xi}{1-\xi}\right)^2 \left[ \frac{1-\xi}{(1-\lambda\xi)^2}
\right]_+ \\ \nonumber
&+&  2\ \frac{\xi}{(1-\xi)_+}\ \frac{1}{(1-\zeta)_{\oplus}}
\left[ 1-(1-\zeta)\ \frac{1-\lambda \xi}{1-\xi} \right] \\
&+& \left. 2\ \xi \left[ 1-(1-\zeta)\ \frac{1-\lambda \xi}{1-\xi}
\right] \right\} \\  \nonumber
H_2^q(\xi,\zeta,\mu^2_F,\lambda) &=& H_1^q(\xi,\zeta,\mu^2_F,\lambda)
+ \frac{4}{3} {\Bigg\{} \delta(1-\xi) \delta(1-\zeta) K_A  \\
&-& \left. 2 \left(\xi(1-3\lambda)[1-(1-\zeta)\frac{1-\lambda \xi}{1-\xi}]
+(1-\lambda)\right)\right\} \\
H_3^q(\xi,\zeta,\mu^2_F,\lambda) &=& H_1^q(\xi,\zeta,\mu^2_F,\lambda)
\ + 2\ \frac{4}{3}\left\{(1-\xi)[1-(1-\zeta)\frac{1-\lambda\xi}{1-\xi}]
-(1-\lambda\xi)\right\} 
\end{eqnarray}
with $\displaystyle \quad
K_A=\frac{1}{\lambda}(1-\lambda)\ln(1-\lambda) \quad$,
$\displaystyle \quad
P_{qq}^{(0)}(\xi)=\frac{4}{3}\left(\frac{1+\xi^2}{1-\xi}\right)_+ \quad$
and where the distributions are defined by
\begin{equation}
\int_0^1 d\xi\ \frac{f(\xi)}{(1-\xi)_+}=
\int_0^1 d\xi\ \frac{f(\xi)-f(1)}{1-\xi}\ ,\ \ \
\int_{\zeta_{min}}^1 d\zeta\ \frac{f(\zeta)}{(1-\zeta)_{\oplus}}=
\int_{\zeta_{min}}^1 d\zeta\ \frac{f(\zeta)-f(1)}{1-\zeta}
\end{equation}
with $\zeta_{min}= (1-\lambda)\xi / (1-\lambda \xi)$. When integrated
over $\zeta$, these results reduce to the final inclusive results
obtained in \cite{ref4,ref5}, i.e.\
\begin{equation}
H_i^q(\xi,\mu^2_F,\lambda) = \int_{\zeta_{min}}^1 d\zeta
\ H_i^q(\xi,\zeta,\mu^2_F,\lambda)\ \ \ .
\end{equation}
The gluonic NLO ($\overline{\rm{{MS}}}$) coefficient functions $H_i^g$
for heavy quark (charm) production in eq.\ (1), as calculated from the
subprocess $W^+ g \rightarrow c {\bar{s}}$, are given by
\begin{eqnarray} \nonumber
H_i^g(\xi,\zeta,\mu^2_F,\lambda) &=&
\delta(1-\zeta) \left\{
P_{qg}^{(0)}(\xi)\left[\ln\frac{Q^2+m_c^2}{\mu^2_F} +
\ln\frac{(1-\xi)^2}{\xi(1-\lambda \xi)}\right]+\xi(1-\xi)\right\} \\
&+& \left[\frac{1}{(1-\zeta)_{\oplus}}+\frac{1}{\zeta}\right]
P_{qg}^{(0)}(\xi)\ +\ h_i^g(\xi,\zeta,\lambda)
\end{eqnarray}
where
\begin{eqnarray}  \nonumber
h_1^g(\xi,\zeta,\lambda) &=&
-\frac{\xi^2}{\zeta^2}(1-\lambda)(1-2\lambda)
+\frac{2\xi}{\zeta}(1-\lambda)(1-2\lambda \xi)  \\
&+& 2\xi\lambda(1-\lambda\xi)-1 \\ \nonumber
h_2^g(\xi,\zeta,\lambda) &=&
\frac{\xi^2}{\zeta^2}(1-\lambda)(1-6\lambda+6\lambda^2) + \frac{6\lambda
\xi}{\zeta}(1-\lambda)(1-2\lambda\xi)  \\
&+& \lambda[6\lambda\xi(1-\lambda\xi)-1] \\
h_3^g(\xi,\zeta,\lambda) &=&
\frac{\xi^2}{\zeta^2}(1-\lambda)(1-2\lambda)-\frac{2}{\zeta}
[P_{qg}^{(0)}(\xi)+\xi(1-\xi-\lambda \xi)(1-\lambda)]
\end{eqnarray}
with $\displaystyle \quad
P_{qg}^{(0)}(\xi)=\frac{1}{2}[\xi^2+(1-\xi)^2] \quad$ and the $\oplus$
distribution is defined in (A4). When integrated
over $\zeta$, these results reduce to the final inclusive results
obtained in \cite{ref4,ref5}, i.e.\
\begin{equation}
H_i^g(\xi,\mu^2_F,\lambda) = \int_{\zeta_{min}}^1 d\zeta
\ H_i^g(\xi,\zeta,\mu^2_F,\lambda)\ \ \ .
\end{equation}
\newpage
%

\newpage
%
%
\newpage
\section*{Figure Captions}
\begin{description}
\item[Fig.\ 1] The predicted $\xi s(\xi,z,Q^2)_{eff}$ in eqs.\ (4) and (5)
as calculated in LO [NLO] using the dynamical (GRV) parton distributions of
ref.\ \cite{ref10} and the charm fragmentation functions in eq.\ (2)
corresponding to $\varepsilon_c=0.20\ [0.06]$. The individual NLO quark--
and gluon--initiated contributions derive in an obvious way from eq.\ (1).
The charm mass has been (arbitrarily) fixed in LO and NLO, $m_c=1.5
\ {\rm{GeV}}$. The shaded band represents the 'data' as extracted from 
\cite{ref7} via eq.\ (3).  
\item[Fig.\ 2] Same as fig.\ 1 but using the LO CTEQ4L and the 
NLO ($\overline{\rm{{MS}}}$) CTEQ4M parton densities \cite{ref11}.
\item[Fig.\ 3] LO and total NLO predictions using different values for
$m_c$ (thin curves). The LO and NLO results for $m_c=1.5\ {\rm{GeV}}$ 
(thick curves) are the same as the corresponding ones in fig.\ 1.
The predictions are based on the GRV densities \cite{ref10}. The shaded 
experimental 'data' band corresponds, however, to $m_c=1.3\ {\rm{GeV}}$
as utilized in LO in \cite{ref1,ref7} and is the same as in fig.\ 1.
\item[Fig.\ 4] Strange sea densities in LO and NLO at 
$Q^2=4\ {\rm{GeV}}^2$ and $Q^2=22\ {\rm{GeV}}^2$ according to CCFR 
\cite{ref1} and CTEQ4 \cite{ref11}. The GRV strange densities \cite{ref10}
refer to \underline{absolute} QCD predictions since they have been 
generated purely radiatively from a vanishing input 
$s(x,\mu^2_{LO,NLO})=0$. The NLO CCFR strange density corresponds to
the choice of the factorization scale $Q^2+m_c^2$ in Table 5 of 
ref.\ \cite{ref1}. 
The actual $x$--ranges covered by the CCFR
data \cite{ref7} at these fixed values of $Q^2$ are indicated. 
\end{description}
\newpage
\pagestyle{empty}

\begin{figure}
\vspace*{-2cm}

\hspace*{-3.5cm}
\epsfig{figure=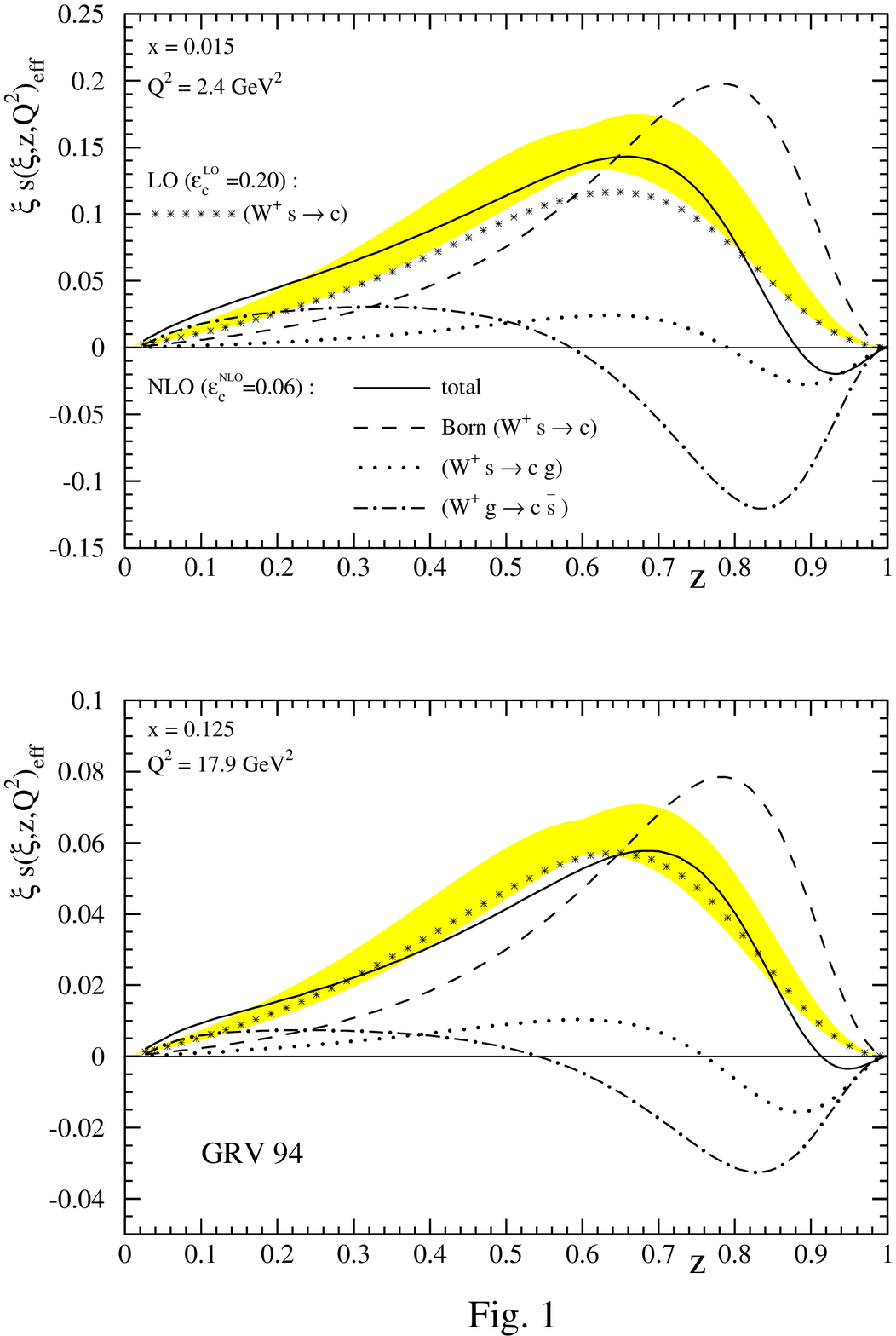,width=20cm}
\end{figure}
\newpage 
  
\begin{figure}
\vspace*{-2cm}

\hspace*{-3.5cm}
\epsfig{figure=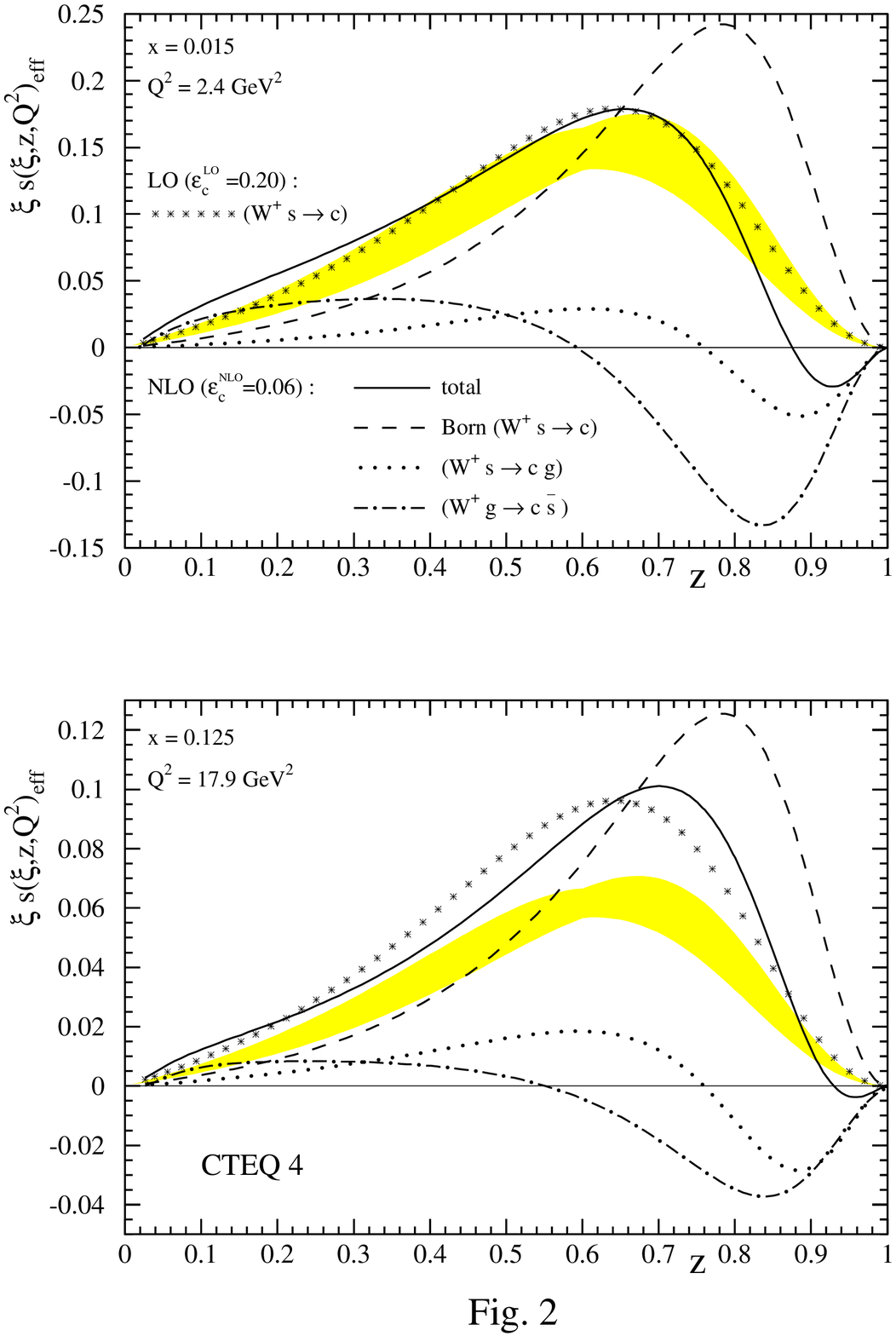,width=20cm}
\end{figure}
\newpage
   
\begin{figure}
\vspace*{-2cm}

\hspace*{-3.5cm}
\epsfig{figure=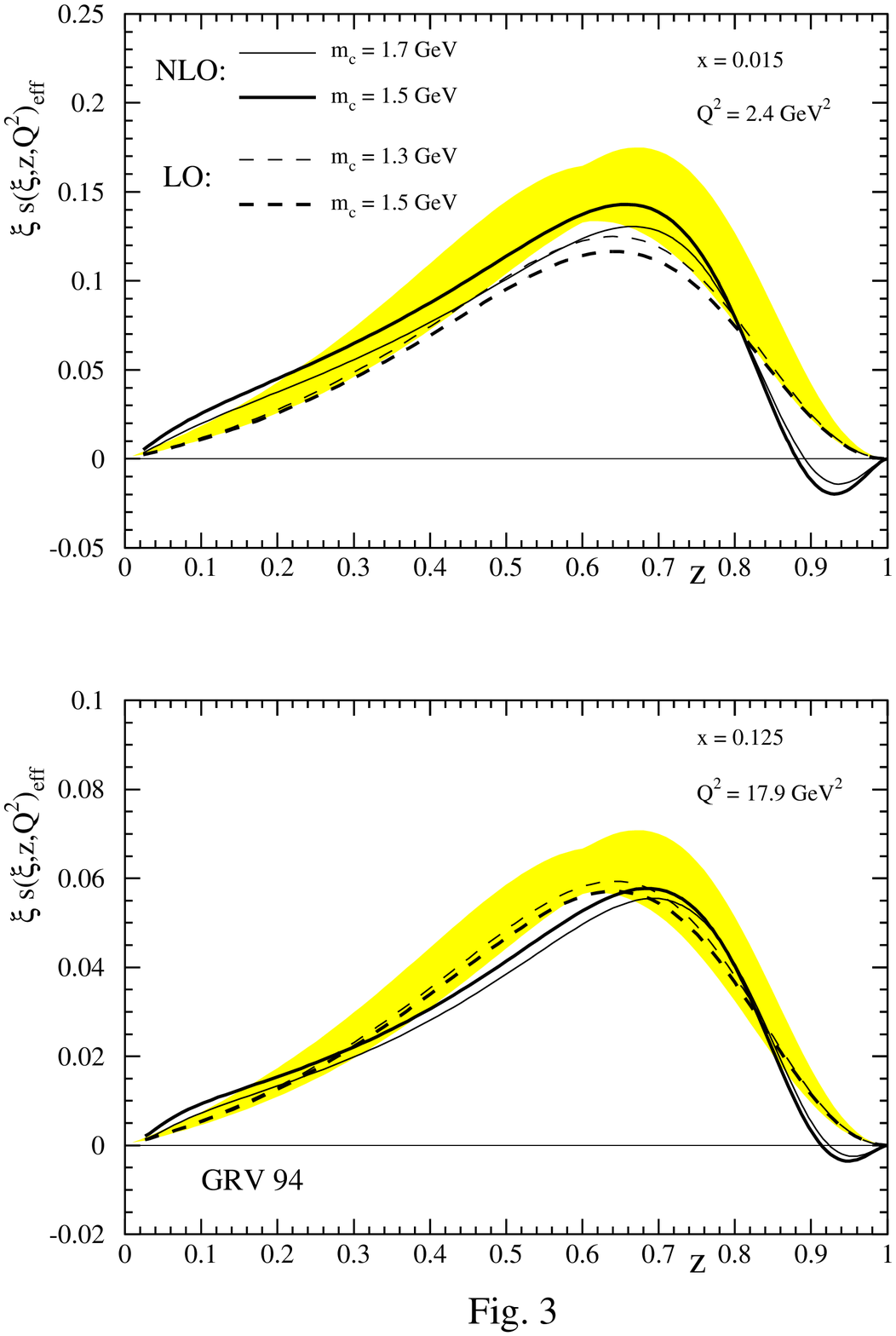,width=20cm}
\end{figure}
\newpage
   
\begin{figure}
\vspace*{-2cm}

\hspace*{-3.5cm}
\epsfig{figure=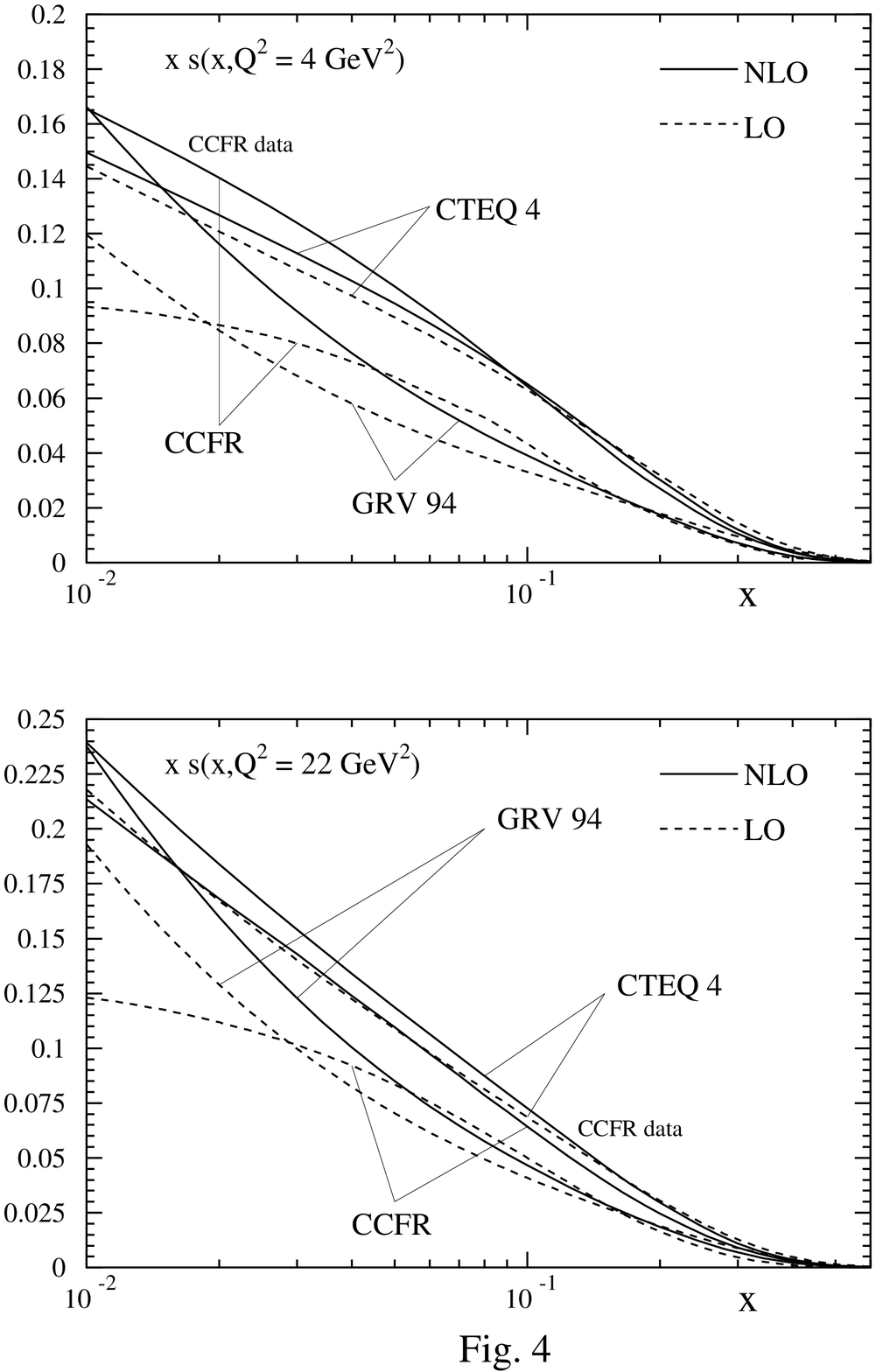,width=20cm}
\end{figure}

\end{document}